\documentclass[12pt]{article}
\usepackage{graphicx}
\usepackage{mathrsfs}
\usepackage{amsmath}
\usepackage{amssymb}
\usepackage{cite}
\newcommand{\citeu}[1]{$^{\mbox{\protect \scriptsize \cite{#1}}}$}

\begin{document}

\renewcommand{\thefootnote}{\fnsymbol{footnote}}
\begin{titlepage}

\vspace{10mm}
\begin{center}
{\Large\bf Interacting Theory of Chiral Bosons and Gauge Fields on Noncommutative Extended Minkowski Spacetime}
\end{center}

\vspace{16mm}
{\large MIAO Yan-Gang${}^{1,2,3,}$\footnote{E-mail: miaoyg@nankai.edu.cn}
and ZHAO Ying-Jie${}^{1,}$\footnote{ E-mail: xiangyabaozhang@mail.nankai.edu.cn}

\vspace{6mm}
${}^{1}${\normalsize  School of Physics, Nankai University, Tianjin 300071, China}

${}^{2}${\normalsize  Kavli Institute for Theoretical Physics China, CAS, Beijing 100190, China}

${}^{3}${\normalsize  The Abdus Salam International Centre for Theoretical Physics,
Strada Costiera 11, 34014 Trieste, Italy}}

\vspace{30mm}
\setlength{\parindent}{0pt}

\textbf{\textsl{Abstract}}
\quad Several interacting models of chiral bosons and gauge fields are investigated
on the noncommutative extended Minkowski spacetime which was recently proposed from a new point of view of disposing noncommutativity.
The models include the bosonized chiral Schwinger model, the generalized chiral Schwinger model (GCSM)
and its gauge invariant formulation. We establish the Lagrangian theories of the models, and then derive the Hamilton's equations in accordance with the
Dirac's method and solve the equations of motion,
and further analyze the self-duality of the Lagrangian theories in terms of the parent action approach.

\vskip 10pt
\textbf{PACS numbers}: 02.40.Gh, 11.10.Nx, 11.10.Kk

\vskip 10pt
\textbf{Key words}: $\kappa$-Minkowski spacetime, noncommutativity, chiral Schwinger model

\end{titlepage}

\newpage
\renewcommand{\thefootnote}{\arabic{footnote}}
\setcounter{footnote}{0}
\setcounter{page}{2}

\section{Introduction}
Physics based on the noncommutative spacetime,\citeu{s1} such as the noncommutative field theory (NCFT),\citeu{s2}
has attracted much attention recently. In mathematics the noncommutative geometry\citeu{s3} has provided a solid basis for the study of physics
related to the noncommutative spacetime.
In light of the Hopf-algebraic method, one can classify the spacetime noncommutativity into the three types,
{\em i.e.} the canonical, Lie-algebraic and quadratic
noncommutativity, respectively. In the three types of noncommutative spacetimes, the ${\kappa}$-deformed Minkowski spacetime\citeu{s4}
as a specific case of the Lie-algebraic type is of particular interest which sometimes implies the foundation of
the Doubly Special Relativity.\citeu{s5}

Usually the noncommutativity can be described by the way of Weyl operators or for the sake of practical
applications by the way of normal functions with a suitable definition of star-products.
As stated in Ref.\cite{s2}, the noncommutativity of spacetimes may be encoded through ordinary products in the
noncommutative $C^{\ast}$-algebra of Weyl operators, or equivalently through the deformation of the product of the
commutative $C^{\ast}$-algebra of functions to a noncommutative star-product. For instance, on the canonical noncommutative spacetime
the star-product is merely the Moyal-product,\citeu{s6} while on the ${\kappa}$-deformed Minkowski spacetime
the star-product requires a more complicated formula.\citeu{s7}

Quite different from the usual scheme mentioned above, one of the present authors proposed\citeu{s8}
a new point of view to dispose the noncommutativity of the ${\kappa}$-deformed Minkowski spacetime.
The motivation is to deal with the ${\kappa}$-deformed Minkowski spacetime in some sense in the same way as the Minkowski spacetime.
By introducing a well-defined proper time from the $\kappa$-deformed Minkowski spacetime that corresponds to the standard basis,
we encode enough information of noncommutativity of the ${\kappa}$-Minkowski spacetime
to a commutative spacetime, and then set up a noncommutative extension of the Minkowski spacetime.
This extended Minkowski spacetime is as commutative as the Minkowski spacetime, but it contains noncommutativity already.
Therefore, one can somehow investigate the noncommutative field theories defined on the ${\kappa}$-deformed Minkowski spacetime by
following the way of the ordinary (commutative) field theories on the noncommutative extension of the Minkowski spacetime,
and thus depict the noncommutativity within the framework of this commutative spacetime.
With this simplified treatment to the noncommutativity of the ${\kappa}$-Minkowski spacetime,
we unveil the fuzziness in the temporal dimension and build noncommutative chiral boson models in Ref.\cite{s8}.

In this paper we discuss the interacting theories of chiral bosons and gauge
fields on the noncommutative extended Minkowski spacetime, which in fact enlarges the scope of Ref.\cite{s8} where only
the free theory of chiral bosons is involved in. Incidentally, in order to avoid repetition we here omit the historical background of the
interacting theories of chiral bosons and gauge fields because the relative context on the chiral boson and its self-duality
has been demonstrated in Ref.\cite{s8} and the references therein. We just simply mention that the main reason that the chiral bosons and the relative
interacting theories with gauge fields have attracted much attention is that they are exactly solvable and can be dealt with as useful theoretical
laboratories in gauge field theory and string theory.
What we have done in Ref.\cite{s8} is to propose and quantize
the Lagrangian theory of noncommutative chiral bosons and to show its preservation of self-duality. In the following sections we focus on the three
interacting models of chiral bosons and gauge fields, {\em i.e.}
the bosonized chiral Schwinger model,\citeu{s9} the generalized chiral Schwinger model (GCSM)\citeu{s10} and its gauge invariant formulation.\citeu{s11}
In light of the scheme given by Ref.\cite{s8} we at first generalize the (commutative) interacting models to their noncommutative formulations,
and then give the Hamilton's equations for the noncommutative Lagrangians by using the Dirac's method\citeu{s12} and solve the equations of motion,
and at last investigate the duality symmetry of the
noncommutative models in terms of the parent action approach.\citeu{s13}

The arrangement of this paper is as follows. In section 2, we briefly review the proposal of the noncommutative extended
Minkowski spacetime. Afterwards
we deal with the three models mentioned above in sections 3, 4 and 5, respectively. Each section includes three subsections.
At the beginning of each section we write
the noncommutative generalization of one model in the extended framework of the Minkowski spacetime.
In the first subsection we derive the Hamilton's equation of the noncommutative model,
and then in the second subsection we solve the equation of motion and give the spectrum, and finally in the third subsection
we discuss the duality symmetry of the noncommutative model.
We make a conclusion in section 6.

\section{Noncommutative extension of the Minkowski spacetime}
In order for this paper to be self-contained we simply repeat the main context of
the noncommutative extension of the Minkowski spacetime.
The major procedure is to connect a commutative
spacetime with the ${\kappa}$-deformed Minkowski spacetime and then to introduce a well-defined proper time. For the details, see Ref.\cite{s8}.

We start with a usual commutative spacetime whose coordinate and momentum operators satisfy the standard Heisenberg commutation relations,
\begin{equation}
[{\hat{\cal X}}^{\mu},{\hat{\cal X}}^{\nu}]=0, \qquad
[{\hat{\cal X}}^{\mu},{\hat{\cal P}}_{\nu}]=i{\delta}^{\mu}_{\nu}, \qquad
[{\hat{\cal P}}_{\mu},{\hat{\cal P}}_{\nu}]=0,               \label{e2.1}
\end{equation}
where ${\mu},{\nu}=0,1,2,3$.
Next, in accordance with Ref.\cite{s14} we give the connection between the commutative spacetime and the ${\kappa}$-deformed Minkowski spacetime
with coordinate and momentum  operators (${\hat{x}}^{\mu}$, ${\hat{p}}_{\nu}$),
\begin{equation}
{\hat x}^0  =  {\hat{\cal X}}^0-\frac{1}{\kappa}[{\hat{\cal X}}^j,{\hat{\cal P}}_j]_{+}, \qquad
{\hat x}^i  =  {\hat{\cal X}}^i+A{\eta}^{ij}{\hat{\cal P}}_j\exp\left(\frac{2}{\kappa}{\hat{\cal P}}_0\right),\label{e2.2}
\end{equation}
where $[{\hat{\cal O}}_1,{\hat{\cal O}}_2]_{+} \equiv \frac{1}{2}\left({\hat{\cal O}}_1{\hat{\cal O}}_2
+{\hat{\cal O}}_2{\hat{\cal O}}_1\right)$, ${\eta}^{{\mu}{\nu}}\equiv{\rm diag}(1,-1,-1,-1)$, $i, j=1,2,3$,
$A$ is an arbitrary constant, and the noncommutative
parameter ${\kappa}$ with the mass dimension is considered to be real and positive. Considering the Casimir operator of the $\kappa$-deformed
Poincar$\acute{\rm e}$ algebra on the standard basis,\citeu{s4}
\begin{equation}
{\hat{\cal C}}_1  = \left({2\kappa \sinh \frac{{\hat p_0}}{{2\kappa}}}
\right)^2  - {\hat p_i}^2,              \label{e2.3}
\end{equation}
we supplement the relations of momentum operators between the commutative spacetime and the ${\kappa}$-Minkowski spacetime as follows:
\begin{equation}
\hat{p}_0  =  2\kappa\,{\sinh}^{-1}\frac{{\hat{\cal P}}_0}{{2\kappa}},\qquad
\hat{p}_i  =  {\hat{\cal P}}_i.             \label{e2.4}
\end{equation}
By using Eqs.~\eqref{e2.1}, \eqref{e2.2} and \eqref{e2.4} we can obtain the complete algebra of the noncommutative phase space
(${\hat{x}}^{\mu}$, ${\hat{p}}_{\nu}$),
\begin{equation*}
[{\hat x}^0,{\hat x}^j]=\frac{i}{\kappa}{\hat x}^j, \qquad [{\hat x}^i,{\hat x}^j]=0, \qquad
[{\hat p}_{\mu},{\hat p}_{\nu}]=0,\qquad [{\hat x}^i,{\hat p}_j]=i{\delta}^{i}_{j},
\end{equation*}
\begin{equation}
[{\hat x}^0,{\hat p}_0]=i\left(\cosh\frac{\hat{p}_0}{{2\kappa}}\right)^{-1},\qquad
[{\hat x}^0,{\hat p}_i]=-\frac{i}{\kappa}{\hat p}_i,\qquad
[{\hat x}^i,{\hat p}_0]=0, \label{e2.5}
\end{equation}
and can further verify that this algebra satisfies the Jacobi identity, which means that we find a consistent relationship
between the commutative spacetime and the ${\kappa}$-Minkowski spacetime, {\em i.e.} Eqs.~\eqref{e2.2} and \eqref{e2.4}.
Moreover, such a relationship makes the above Casimir operator have the usual formula as expected,
\begin{equation}
{\hat{\cal C}}_{1} =  {{\hat{\cal P}}_0}^2-{{\hat{\cal P}}_i}^2, \label{e2.6}
\end{equation}
which coincides with the standard Heisenberg commutation relations (Eq.~\eqref{e2.1}).

If $\hat p_\mu$ takes the usual forms,
\begin{equation}
\hat{p}_0 = -i\frac{\partial}{{\partial}t},\qquad
\hat{p}_i = -i\frac{\partial}{{\partial}x^i},    \label{e2.7}
\end{equation}
the operator ${\hat{\cal P}}_0$ then reads
\begin{equation}
{\hat{\cal P}}_0=-i{2\kappa}\left(\sin\frac{1}{2\kappa}\frac{{\partial}}{{\partial}t}\right). \label{e2.8}
\end{equation}
We deal with $t$ as the parameter describing the dynamical evolution of fields.
It tends to the ordinary time variable in the limit $\kappa \rightarrow +\infty$, which guarantees the consistency of the choice of the parameter.

We now introduce a proper time $\tau$ by defining the operator
\begin{equation}
{\hat{\cal P}}_0 \equiv -i\frac{{\partial}}{{\partial}\tau}, \label{e2.9}
\end{equation}
and then postulate\footnote{In general, one may postulate $[\hat{\cal P}_0, \tau^m] = -im\tau^{m-1}$,
$m \in \mathbb{N}$, where a different $m$ corresponds to a different function $\tau(t)$ which leads to a different extended Minkowski spacetime.
Here and in Ref.\cite{s8}, we only consider the case $m=1$,  that is the reason why this case, i.e. Eq.~\eqref{e2.10} is called
the postulation of the operator linearization. We note that each case for a definite $m$ corresponds to one extended Minkowski spacetime
and the number of ways to
map the $\kappa$-deformed Minkowski spacetime to an extended Minkowski spacetime is infinite.
In the limit $\kappa\rightarrow\infty$,
such a mapping is unique, which shows the consistency of the mapping.
For the details on the clarifications of the mapping
and of the reasonability of the choice of $m=1$, see Ref.\cite{s8}. }
 a linear realization or representation of operator ${\hat{\cal P}}_0$ by using Eqs.~\eqref{e2.8} and \eqref{e2.9},
\begin{equation}
2\kappa\left(\sin\frac{1}{2\kappa}\frac{d}{\mathrm{d}t}\right)\tau=1. \label{e2.10}
\end{equation}
The solution of this differential equation gives a well-defined proper time
\begin{equation}
\tau=t+\sum_{n=0}^{+\infty}c_{-n}\exp(-2{\kappa}n{\pi}t), \label{e2.11}
\end{equation}
where $n$ takes zero and positive integers and the coefficient $c_{-n}$ is an arbitrary real constant which presents a kind
of temporal fuzziness compatible with the ${\kappa}$-Minkowski spacetime.\citeu{s4}\citeu{s8} In the limit $\kappa \rightarrow +\infty$,
the proper time turns back to the ordinary time variable.

We thus build the noncommutative extension of the Minkowski spacetime $(\tau, x^i)$ to which the
information of noncommutativity has been encoded through the proper time. This can be seen clearly when the extended spacetime is transformed into
the coordinates $(t, x^i)$. In addition, we point out that the extended Minkowski spacetime is a special flat spacetime
corresponding to a twisted $t$-parameter, which is obvious from its metric
\begin{equation}
g_{00}={\dot \tau}^2=\left[1-2{\kappa}{\pi}\sum_{n=0}^{+\infty}nc_{-n}\exp(-2{\kappa}n{\pi}t)\right]^2,
\qquad g_{11}=g_{22}=g_{33}=-1.\label{e2.12}
\end{equation}
Because it is based on the standard Heisenberg commutation relations (Eq.\eqref{e2.1}) the extended Minkowski spacetime is, as expected, commutative.
We can utilize this merit to construct noncommutative models in the commutative framework. That is, by simply considering the Lorentz
invariance on the extended Minkowski spacetime we can naturally write the Lagrangian of a model that contains noncommutative effects.
The concrete procedure is as follows:\footnote{In the following three sections we focus our discussions on the $(1+1)$-dimensional spacetimes.}
the Lagrangian of a noncommutative model is given
by the requirement of the Lorentz invariance on the extended Minkowski spacetime spanned by the
coordinates $(\tau, x)$, and through the coordinate transformation Eq.~\eqref{e2.11},
it is then converted into its $(t, x)$-coordinate formulation with explicit noncommutativity.
As a result, we establish the Lagrangian theory of the noncommutative model in the extended framework of the Minkowski spacetime.
This procedure will be applied to the three models\citeu{s9,s10,s11} in the following three sections, respectively.

\section{Bosonized chiral Schwinger model}
Considering the Lorentz invariance on the extended Minkowski spacetime\footnote{The metric is the same as that of the Minkowski spacetime, {\em i.e.},
${\eta}^{\mu \nu}={\rm diag}(1,-1)$.} $(\tau, x)$,
we give the action of the bosonized chiral Schwinger model,\citeu{s9}
\begin{eqnarray}
S_{1}&=&\int \mathrm{d}{\tau}\mathrm{d}x \bigg[\frac{\partial{\phi}}{\partial{\tau}}\frac{\partial{\phi}}{\partial{x}}
- \left(\frac{\partial{\phi}}{\partial{x}}\right)^2
+ 2e\frac{\partial{\phi}}{\partial{x}}\left( {A_0  - A_1 } \right)
 - \frac{1}{2}e^2 \left( {A_0  - A_1 } \right)^2\bigg. \nonumber \\
& &\bigg.+ \frac{1}{2}e^2 a\,{\eta}^{\mu \nu }A_\mu  A_{\nu}- \frac{1}{4}{\eta}^{\mu \rho }{\eta}^{\nu \sigma}F_{\mu\nu}F_{\rho\sigma}\bigg],
\label{e3.1}
\end{eqnarray}
where $\phi$ is a chiral boson field; $A_{\mu}$ is a gauge field and $F_{\mu\nu}={\partial}_{\mu}A_{\nu}-{\partial}_{\nu}A_{\mu}$ its field strength;
${\eta}^{\mu \nu}={\rm diag}(1,-1)$ is the flat metric of the extended Minkowski spacetime $(\tau, x)$;
$e$ is the electronic charge and $a$ is a real parameter\footnote{In general $a > 1$, see also Ref.\cite{s9} for the details.}
which presents a kind of ambiguity of bosonization. After making the coordinate transformation, we
obtain the action written in terms of the coordinates $(t,x)$,
\begin{eqnarray}
S_{1}&=&\int \mathrm{d}t \mathrm{d}x \sqrt{-g}\bigg\{\frac{1}{{\dot \tau }}\frac{\partial{\phi}}{\partial{t}}\frac{\partial{\phi}}{\partial{x}}
- \left(\frac{\partial{\phi}}{\partial{x}}\right)^2 + 2e\frac{\partial{\phi}}{\partial{x}}\left( {A_0  - A_1 } \right)
- \frac{1}{2}e^2 \left( {A_0  - A_1 } \right)^2\bigg. \nonumber \\
& &\bigg.+ \frac{1}{2}e^2 a\left[\left(A_0\right)^2-  \left(A_1\right)^2\right]
+\frac{1}{2}\left({\frac{1}{{\dot\tau}}\frac{{\partial A_1}}{{\partial t}}-\frac{{\partial A_0}}{{\partial x}}} \right)^2\bigg\},\label{e3.2}
\end{eqnarray}
where $\sqrt{-g}$ is the Jacobian and also the nontrivial measure of the flat spacetime (see Eq.~\eqref{e2.12}) connected with the
${\kappa}$-Minkowski spacetime. Note that $\sqrt{-g}=\vert{\dot \tau}\vert$ in general. Here we just focus on the case ${\dot \tau} > 0$.
As to ${\dot \tau} < 0$, we can make a similar discussion. For the details, see Ref.\cite{s8}.
Therefore the Lagrangian takes the form,
\begin{eqnarray}
\mathcal{L}_1 & = & \dot \phi \phi ' - \sqrt { - g} \left( {\phi '} \right)^2  + \sqrt { - g} \left\{ {2e\phi '\left( {A_0  - A_1 } \right)
- \frac{1}{2}e^2 \left( {A_0  - A_1 } \right)^2  + \frac{1}{2}e^2 a\left[\left(A_0\right)^2-  \left(A_1\right)^2\right]  } \right\}  \nonumber \\
& &+ \frac{1}{{2\sqrt { - g} }}\left( {\dot A_1  - \sqrt { - g}  A_0 '} \right)^2,\label{e3.3}
\end{eqnarray}
where a dot and a prime stand for derivatives with respect to time\footnote{For the sake of convenience
in description, here the {\em time},  different from the ordinary time variable, stands only for a {\em t-parameter}
in sections 3, 4 and 5.}
$t$ and space $x$, respectively. This is the noncommutative generalization of the bosonized chiral Schwinger model, it contains the noncommutativity
through the proper time $\tau$ with the finite noncommutative parameter $\kappa$.
In the limit $\kappa\rightarrow +\infty$, $\sqrt { - g}  = {\dot \tau } =1$, the Lagrangian turns back to its ordinary form on the
Minkowski spacetime.

\subsection{Equation of motion}
As the bosonized chiral Schwinger model is a constrained system with second-class constraints, we therefore derive the Hamilton's equations
for the chiral boson and gauge field by using the
Dirac's method.\citeu{s12} First we define the momenta conjugate to $\phi$, $A_0$ and $A_1$, respectively,
\begin{equation}
\pi_\phi   \equiv  \frac{{\partial \mathcal{L}_1}}{{\partial \dot \phi}} =   \phi ', \qquad
\pi ^0   \equiv   \frac{{\partial \mathcal{L}_1}}{{\partial {\dot A_0}}} \approx 0,\qquad
\pi ^1   \equiv    \frac{{\partial \mathcal{L}_1}}{{\partial \dot A_1}} = \frac{1}{{\sqrt { - g}}}\left({\dot A_1  - \sqrt {- g}  A_0 '} \right),
\label{e3.4}
\end{equation}
and then give the Hamiltonian through the Legendre transformation,
\begin{eqnarray}
\mathcal{H}_1 &\equiv& \pi _\phi  \dot \phi  + \pi ^\mu  \dot A_\mu   - \mathcal{L}_1  \nonumber \\
&= &\sqrt { - g} \left\{ \frac{1}{2}\left( {\pi ^1 } \right)^2 + \pi ^1 \partial _1 A_0  + \left( {\phi '} \right)^2
- 2e\phi '\left( {A_0  - A_1 } \right)\right.  \nonumber \\
& & \left.+ \frac{1}{2}e^2 \left( {A_0  - A_1 } \right)^2- \frac{1}{2}e^2 a\left[\left(A_0\right)^2-  \left(A_1\right)^2\right] \right\}.\label{e3.5}
\end{eqnarray}

The definition of momenta (Eq.~\eqref{e3.4}) provides in fact two primary constraints,
\begin{equation}
\Omega _1  \equiv \pi ^0  \approx 0, \qquad  \Omega _2  \equiv \pi _\phi   - \phi ' \approx 0,\label{e3.6}
\end{equation}
where ``$\approx$'' stands for the Dirac's weak equality. In light of the consistency of constraints under the time evolution,
we deduce one secondary constraint from $\Omega _1$,
\begin{equation}
\Omega _3  \equiv  \partial _1 \pi ^1  + 2e\phi ' + e^2 \left[ {\left( {a - 1} \right)A{}_0 + A_1 } \right] \approx 0,\label{e3.7}
\end{equation}
but no further constraints from $\Omega _2$ and $\Omega _3$. Therefore the three constraints constitute a complete set with the
non-vanishing equal-time Poisson brackets as follows:
\begin{eqnarray}
\left\{{\Omega_1 \left( x \right),\Omega _3 \left( y \right)} \right\}_{\rm PB} &=& -e^2\left( {a - 1}\right)\delta \left( {x - y}\right),\nonumber \\
\left\{{\Omega_2 \left( x \right),\Omega _2 \left( y \right)} \right\}_{\rm PB} &=& -2\partial _x \delta \left( {x - y} \right),\nonumber \\
\left\{{\Omega_2 \left( x \right),\Omega _3 \left( y \right)} \right\}_{\rm PB} &=& 2e\partial_x \delta \left( {x - y} \right),\nonumber \\
\left\{{\Omega_3 \left( x \right),\Omega _3 \left( y \right)} \right\}_{\rm PB} &=& -2e^2 \partial _x\delta \left( {x - y} \right).\label{e3.8}
\end{eqnarray}
Calculating the inverse elements of Poisson brackets and utilizing the definition of Dirac brackets,\citeu{s12}
we finally obtain the non-vanishing equal-time Dirac brackets for the  chiral boson and gauge field,
\begin{eqnarray}
\left\{ {\phi \left( x \right),\phi \left( y \right)} \right\}_{\rm DB}  &=&  - \frac{1}{2} \varepsilon \left( {x - y} \right),\nonumber \\
\left\{ {\phi \left( x \right),\pi _\phi  \left( y \right)} \right\}_{\rm DB}  &=& \frac{1}{2}\delta \left( {x - y} \right), \nonumber \\
\left\{ {A_1 \left( x \right),\pi ^1 \left( y \right)} \right\}_{\rm DB} & =& \delta \left( {x - y} \right),\label{e3.9}
\end{eqnarray}
where $\varepsilon(x)$ is the step function, $d{\varepsilon(x)}/{\mathrm{d}x} = \delta \left( x \right)$.

When the Dirac weak constraints become strong conditions, we write the reduced Hamiltonian in terms of the independent variables of phase space,
{\em i.e.}, $\phi$, $A_1$ and $\pi^1$,
\begin{eqnarray}
\mathcal{H}^{\rm r}_1 &=& \sqrt { - g} \bigg[ \frac{1}{2}\left( {\pi ^1 } \right)^2
+ \frac{1}{{2e^2 \left( {a - 1}\right)}}{\left( {\partial _1 \pi ^1 } \right)^2}
+\frac{{e^2 a^2}}{{2\left({a-1}\right)}}\left({A_1}\right)^2
+ \frac{{a + 1}}{{a - 1}}\left( \phi '\right)^2  \bigg. \nonumber \\
& & \bigg.+\frac{1}{{a-1}}{A_1 \partial_1\pi^1}+\frac{2}{{e\left( {a - 1} \right)}}{\phi '\partial_1\pi ^1 }
+ \frac{{2ea}}{{a - 1}}\phi '  A_1  \bigg],\label{e3.10}
\end{eqnarray}
and then get the Hamilton's equations with the formula,
$\dot F(x)= \int {dy} \left\{ {F \left( x \right),\mathcal{H}^{\rm r}_1 \left( y \right)} \right\}_{\rm DB}$,
\begin{eqnarray}
\dot \phi
&=& \sqrt { - g} \left[ \frac{{a + 1}}{{a - 1}}\phi ' + \frac{1}{{e\left( {a - 1} \right)}}{\partial _1 \pi ^1 }
+ \frac{{ea}}{{a - 1}}A_1 \right], \nonumber  \\
\dot A_1
&=& \sqrt { - g}\left[-\frac{1}{a-1}{\partial _1 A_1}+\pi ^1-\frac{1}{e^2(a-1)}{\partial}_1{\partial}_1\pi^1-\frac{2}{e(a-1)}\phi ''\right],\nonumber \\
{\dot \pi ^1}
&=&\frac{{\sqrt { - g} }}{{a - 1}}\left[ { - {\partial _1}{\pi ^1} - 2ea{\phi '} - {e^2}{a^2}{A_1}} \right].\label{e3.11}
\end{eqnarray}
As to the equations of motion for the other three phase space variables, we can easily derive from the constraints (Eqs.~\eqref{e3.6} and \eqref{e3.7})
with the replacement of the Dirac weak equality by the strong one.

\subsection{Solution}
By eliminating the momenta from the above Hamilton's equations, we obtain the Euler-Lagrange equations for $\phi$, $A_0$ and $A_1$, respectively,
\begin{eqnarray}
\partial _1 \left\{ { \partial _0 \phi- \sqrt { - g} \left[\partial _1 \phi -  e\left( {A_0 - A_1 } \right)\right]} \right\} &=& 0,  \nonumber \\
\partial _1 \left(\frac{1} {\sqrt { - g} }\partial _0 A_1 -  \partial _1 A_0  \right) +   2e\partial _1 \phi
+ e^2  \left[ \left( a-1 \right){A_0  } +  A_1\right] &=& 0, \nonumber \\
\frac{1}{\sqrt { - g}}{\partial _0}\left(\frac{1}{\sqrt { - g}} {\partial _0}{A_1}- {\partial _1}{A_0}\right)+2e{\partial _1}\phi
-{e^2} \left[{A_0} - \left(a+1\right){A_1}  \right] &=& 0.\label{e3.12}
\end{eqnarray}
After comparing them with that of Ref.\cite{s9} and doing a tedious calculation, we at last have the solutions as follows:
\begin{eqnarray}
\phi  &=& \sigma  - h, \nonumber \\
A_0
&=&  - \frac{1}{{ea}}\left[ {\left( {  \frac{1}{{\sqrt { - g} }}{\partial _0 } + \partial _1 } \right)\left( {\sigma  - h} \right)
- a\partial _1 \sigma } \right],  \nonumber \\
A_1
&=&  - \frac{1}{{ea}}\left[ {\left( { \frac{1}{{\sqrt { - g} }} {\partial _0 }+ \partial _1 } \right)\left( {\sigma  - h} \right)
- \frac{a}{{\sqrt { - g} }}\partial _0 \sigma } \right],  \label{e3.13}
\end{eqnarray}
where $h$ and $\sigma$ are new fields introduced and they satisfy the following equations of motion,
\begin{eqnarray}
\left( {  \frac{1}{{\sqrt { - g} }}{\partial _0 } - \partial _1 } \right)h &=& 0, \label{e3.14} \\
\frac{1}{{\sqrt { - g} }}{\partial _0}\left( {\frac{1}{{\sqrt { - g} }}{\partial _0}\sigma } \right) + {\partial ^1}{\partial _1}\sigma
+ \frac{{{e^2}{a^2}}}{{a - 1}}\sigma  &=& 0.   \label{e3.15}
\end{eqnarray}
Eq.~\eqref{e3.14} is just the noncommutative left-moving chiral boson\citeu{s8} while Eq.~\eqref{e3.15} the noncommutative Klein-Gordon equation
describing a free and massive scalar boson on the extended Minkowski spacetime.
Moreover, by using Eqs. \eqref{e3.9}, \eqref{e3.11} and \eqref{e3.13}, we work out the equal-time Dirac brackets for the newly introduced fields,
\begin{eqnarray}
{\left\{ {h\left( x \right),h\left( y \right)} \right\}_{\rm DB}} &=& - \frac{1}{2} \varepsilon \left( {x - y} \right),\nonumber \\
{\left\{ {h\left( x \right),\dot h\left( y \right)} \right\}_{\rm DB}} &=& \frac{{\sqrt { - g} }}{2}\delta \left( {x - y} \right),\nonumber \\
{\left\{ {\sigma \left( x \right),\sigma \left( y \right)} \right\}_{\rm DB}} &=& 0, \nonumber  \\
{\left\{ {\sigma \left( x \right),\dot \sigma \left( y \right)} \right\}_{\rm DB}} &=& \frac{{\sqrt { - g} }}{{a - 1}}\delta \left( {x - y} \right).
\label{e3.16}
\end{eqnarray}

In consequence we solve completely the noncommutative generalization of the bosonized chiral Schwinger model which is depicted by the Lagrangian
Eq.~\eqref{e3.3} and find that the spectrum of the model includes a chiral boson $h(x)$ with the left-chirality and a massive scalar field $\sigma(x)$
with the mass $m^2={e^2}{a^2}/{(a - 1)}$ in the framework of the extended Minkowski spacetime.
In addition, we note that the equations of motion (Eqs.~\eqref{e3.14} and \eqref{e3.15}) and the Dirac brackets (Eq.~\eqref{e3.16})
take their usual forms on the extended Minkowski spacetime
with the $(\tau, x)$ coordinates. This shows the consistency of our generalization and also provides a simple way to solve Eq.~\eqref{e3.15}.
As was dealt with to the noncommutative chiral boson,\citeu{s8} one can solve the noncommutative Klein-Gordon equation in terms of the well-defined
light-cone coordinates which contain the noncommutativity already.

\subsection{Duality symmetry}
We investigate the duality symmetry of the noncommutative generalization of the bosonized chiral Schwinger model in accordance with
the parent action approach.\citeu{s13} Here we just simply summarize the procedure of this approach. For the details on its historical background
and its significance in field theory and string theory, see Ref.\cite{s13} and the
references therein. The approach mainly includes the following three steps:
\begin{itemize}
\item
to introduce auxiliary fields and then to construct a parent or master
action based on a source action;
\item
to make the variation of the parent
action with respect to each auxiliary field, to solve one auxiliary field
in terms of other fields and then to substitute the solution into the parent
action;
\item
after step two, one can obtain different forms of an action. These forms are, of
course, equivalent classically, and the relation between them is usually
referred to duality. If the resulting forms are same, their relation is called self-duality.
\end{itemize}

We introduce two auxiliary vector fields $F_\mu$ and $G^\mu$, and write the following parent action corresponding to Eq.~\eqref{e3.2},
\begin{eqnarray}
S_1^{\rm p} &=& \int {\mathrm{d}t\mathrm{d}x} \bigg\{ F_0 F_1- \sqrt { - g} {{\left( {F_1} \right)}^2}
+ \frac{1}{{2\sqrt { - g} }}{{\left( {{\partial _0}{A_1}
- \sqrt { - g} {\partial _1}{A_0}} \right)}^2} \bigg. \nonumber \\
& &+ \sqrt { - g} \left[ {2e F_1\left( {{A_0} - {A_1}} \right) - \frac{1}{2}{e^2}{{\left( {{A_0} - {A_1}} \right)}^2}
+ \frac{1}{2}{e^2}a\left[({A_0})^2-({A_1})^2\right]} \right] \nonumber \\
& &\bigg.+G^{\mu}\left(F_{\mu}-{\partial}_{\mu}\phi\right)\bigg\}.\label{e3.17}
\end{eqnarray}
The variation of Eq.~\eqref{e3.17} with respect to $G^\mu$ gives $F_\mu = {\partial _\mu }\phi$, which simply yields the equivalence between
Eq.~\eqref{e3.2} and Eq.~\eqref{e3.17}.
However, making the variation of Eq.~\eqref{e3.17} with respect to $F_\mu$  we have
\begin{eqnarray}
{F_0} &=& - 2\sqrt { - g} {G_0} + {G_1} - 2e\sqrt { - g} \left( {{A_0} - {A_1}} \right), \nonumber \\
{F_1} &=&  - {G_0}.\label{e3.18}
\end{eqnarray}
Substituting Eq.~\eqref{e3.18} into the action Eq.~\eqref{e3.17}, we obtain a kind of dual versions for the action $S_1$ (Eq.~\eqref{e3.2}),
\begin{eqnarray}
{\tilde S}_1^{\rm dual}&=& \int {\mathrm{d}t\mathrm{d}x} \bigg\{G_0 G_1- \sqrt { - g} {{\left( {G_0} \right)}^2}
+ \frac{1}{{2\sqrt { - g} }}{{\left( {{\partial _0}{A_1} - \sqrt { - g} {\partial _1}{A_0}} \right)}^2} \bigg. \nonumber \\
& & + \sqrt { - g} \left[ -2e G_0\left( {{A_0} - {A_1}} \right) - \frac{1}{2}{e^2}{{\left( {{A_0} - {A_1}} \right)}^2}
+ \frac{1}{2}{e^2}a\left[({A_0})^2-({A_1})^2\right] \right] \nonumber \\
& & \bigg. +\phi\partial _\mu G^\mu \bigg\},\label{e3.19}
\end{eqnarray}
Finally, we make the variation of Eq.~\eqref{e3.19} with respect to $\phi$ and have ${\partial _\mu }{G^\mu } = 0$, whose solution is
\begin{equation}
{G^\mu }\left( \varphi  \right) = -{\epsilon ^{\mu \nu }}{\partial _\nu }\varphi \equiv -{\epsilon ^{\mu \nu }}F_{\nu }(\varphi),\label{e3.20}
\end{equation}
where $\epsilon ^{01}=-\epsilon ^{10}=1$, and $\varphi(x)$ is an arbitrary scalar field. Substituting Eq.~\eqref{e3.20} into Eq.~\eqref{e3.19}
we work out the dual action in terms of $\varphi$,
\begin{eqnarray}
S_1^{\rm dual} &=& \int {\mathrm{d}t\mathrm{d}x} \bigg\{{\dot \varphi }{\varphi '} - \sqrt { - g} {{\left( {\varphi '} \right)}^2}
+ \frac{1}{{2\sqrt { - g} }}{{\left( {{\partial _0}{A_1} - \sqrt { - g} {\partial _1}{A_0}} \right)}^2}\bigg. \nonumber \\
& & \bigg. + \sqrt { - g} \left[ {2e {\varphi '} \left( {{A_0} - {A_1}} \right)
- \frac{1}{2}{e^2}{{\left( {{A_0} - {A_1}} \right)}^2} + \frac{1}{2}{e^2}a\left[({A_0})^2-({A_1})^2\right]} \right]\bigg\}.\label{e3.21}
\end{eqnarray}
This action has the same form as the original action Eq.~\eqref{e3.2} only with the replacement of $\phi$ by $\varphi$.
As a result, the noncommutative generalization of the bosonized chiral Schwinger model is self-dual with respect
to the anti-dualization of ${G^\mu }\left( \varphi  \right)$ and ${F_\mu }\left( \varphi \right)$ (Eq.~\eqref{e3.20}).

\section{Generalized chiral Schwinger model (GCSM)}
The bosonic action of the GCSM can be written\citeu{s10} as the covariant formulation on the extended Minkowski spacetime with
$(\tau, x)$-coordinates,
\begin{equation}
S_{2} = \int \mathrm{d}\tau \mathrm{d}x\left[ \frac{1}{2}\left( {\partial {}_\mu \phi } \right)\left( {{\partial ^\mu }\phi } \right)
+ e{A_\mu }\left( {{\epsilon ^{\mu \nu }} - r{{\eta}^{\mu \nu }}} \right){\partial_\nu }\phi
+ \frac{1}{2}{e^2}a{A_\mu }{A^\mu } - \frac{1}{4}{F_{\mu \nu }}{F^{\mu \nu }} \right],   \label{e4.1}
\end{equation}
where $\phi$ is an auxiliary scalar field introduced in order to give a local $S_2$,
$r$ is a real parameter interpolating between the vector ($r = 0$) and the chiral ($r = \pm 1$) Schwinger models.
This action can be rewritten in terms of $(t, x)$-coordinates as follows,
\begin{eqnarray}
S_2 &=& \int {\mathrm{d}t \mathrm{d}x} \Bigg\{ \frac{1}{{2\sqrt { - g} }}\left[{{\left( {\frac{{\partial \phi }}{{\partial t}}} \right)}^2}
-{{\left( \sqrt { - g}{\frac{{\partial \phi }}{{\partial x}}} \right)}^2}\right]
+ \frac{1}{{2\sqrt { - g} }}\left( {\frac{{\partial {A_1}}}{{\partial t}}  - \sqrt { - g}  \frac{{\partial {A_0}}}{{\partial x}}} \right)^2 \Bigg.
\nonumber \\
& & \Bigg.- e\left( {r{A_0}+ {A_1}} \right)\frac{{\partial \phi }}{{\partial t}}
+e \sqrt { - g} \left( {{A_0} + r{A_1}} \right)\frac{{\partial \phi }}{{\partial x}}
+ \frac{1}{2}{e^2}a\,\sqrt{-g}\left[({A_0})^2-({A_1})^2\right] \Bigg\}, \label{e4.2}
\end{eqnarray}
where the noncommutativity presents explicitly through the Jacobian $\sqrt { - g}$.
Therefore we obtain the Lagrangian of the noncommutative GCSM,
\begin{eqnarray}
\mathcal{L}_2 &= &\frac{1}{{2\sqrt { - g} }}\left[ (\dot \phi)^2
-{{\left( \sqrt { - g}{\phi '} \right)}^2}\right]
+ \frac{1}{{2\sqrt { - g} }}\left( \dot A_1  - \sqrt { - g}  {A_0 '}\right)^2
\nonumber \\
& & - e\left( {r{A_0}+ {A_1}} \right)\dot \phi
+e \sqrt { - g} \left( {{A_0} + r{A_1}} \right){\phi '}
+ \frac{1}{2}{e^2}a\,\sqrt{-g}\left[({A_0})^2-({A_1})^2\right].\label{e4.3}
\end{eqnarray}
In the limit $\kappa \rightarrow \infty$ the action turns out to be its ordinary form on the Minkowski spacetime,
which shows the consistency of our noncommutative generalization.

\subsection{Equation of motion}
Now we derive the Hamilton's equations from the Lagrangian Eq.~\eqref{e4.3}. By the use of the Dirac's method, we at first
define the momenta conjugate to $\phi$, $A_0$ and $A_1$, respectively,
\begin{equation}
\pi _\phi   \equiv \frac{{\partial \mathcal{L}_2}}{{\partial \dot \phi}}
= \frac{1}{{\sqrt { - g} }}\dot\phi  - e\left( {r{A_0} + {A_1}} \right), \qquad
\pi ^0  \equiv \frac{{\partial \mathcal{L}_2}}{{\partial \dot A_0 }} \approx 0, \qquad
\pi ^1  \equiv \frac{{\partial \mathcal{L}_2}}{{\partial \dot A_1 } } = \frac{1}{{\sqrt { - g} }}\dot A_1 -  {A_0 '},\label{e4.4}
\end{equation}
and then give the Hamiltonian
\begin{eqnarray}
\mathcal{H}_2 &\equiv &\pi ^\mu  \dot A_\mu   + \pi _\phi  \dot \phi  - \mathcal{L}_2 \nonumber \\
  &=&{\sqrt { - g} }\left\{\frac{1}{2}{\left[ {{\pi _\phi } + e\left( {r{A_0} + {A_1}} \right)} \right]^2}
  + \frac{1}{2}{\left( {{\partial _1}\phi } \right)^2} + \frac{1}{2}{\left( {{\pi ^1} + {\partial _1}{A_0}} \right)^2}
  - \frac{1}{2}{\left( {{\partial _1}{A_0}} \right)^2}\right.  \nonumber \\
  & &\left.-  e\left( {{A_0} + r{A_1}} \right)\left( {{\partial _1}\phi } \right) - \frac{1}{2}{e^2}a\left[\left(A_0\right)^2
  -  \left(A_1\right)^2\right]\right\} .\label{e4.5}
\end{eqnarray}

The definition of momenta (Eq.~\eqref{e4.4}) gives us one primary constraint,
\begin{equation}
\Omega _1  \equiv \pi ^0  \approx 0,\label{e4.6}
\end{equation}
and its consistency under the time evolution, $\dot{\Omega }_1\approx 0$, provides one secondary constraint,
\begin{eqnarray}
\Omega _2  \equiv -er\left[ {  {\pi _\phi } + e\left( {r{A_0} + {A_1}} \right)} \right] + e{\partial _1}\phi  + {\partial _1}\pi^1+ e^2a{A_0}
\approx 0.\label{e4.7}
\end{eqnarray}
We have no further constraints from $\Omega _2$.
As a result, the two constraints constitute a complete set with the non-vanishing equal-time Poisson bracket as follows:
\begin{eqnarray}
\left\{{\Omega_1\left(x\right), \Omega_2\left(y\right)}\right\}_{\rm PB}&=&{e^2}\left( {{r^2} - a} \right)\delta \left( {x - y} \right).\label{e4.8}
\end{eqnarray}
From the inverse elements of Poisson brackets and the definition of Dirac brackets,\citeu{s12}
we derive the non-vanishing equal-time Dirac brackets for the chiral boson and gauge field,
\begin{eqnarray}
\left\{ {\phi \left( x \right),\pi _\phi  \left( y \right)} \right\}_{\rm DB}  &= &\delta \left( {x - y} \right), \nonumber\\
\left\{ {A_1 \left( x \right),\pi ^1 \left( y \right)} \right\}_{\rm DB}  &=& \delta \left( {x - y} \right).\label{e4.9}
\end{eqnarray}

When the Dirac weak constraints are replaced by the strong conditions,
we obtain the reduced Hamiltonian in terms of the independent variables of phase space,
{\em i.e.}, $\phi$, $\pi_\phi$, $A_1$ and $\pi^1$,
\begin{eqnarray}
\mathcal{H}^{\rm r}_2 &=& {\sqrt { - g} }\left\{\frac{1}{2}{\left( { e{A_1} + {\pi _\phi }} \right)^2} + \frac{1}{2}{\left( {\phi '} \right)^2}
+ \frac{1}{2}{\left( {{\pi ^1}} \right)^2}- er{A_1} {\phi '}  + \frac{1}{2}{e^2}a{\left( {{A_1}} \right)^2}\right.\nonumber\\
& &\left. - \frac{1}{{2{e^2}\left( {{r^2} - a} \right)}}{\left[ { - er{\pi _\phi } - {e^2}r{A_1} + e\phi ' + {\partial _1}{\pi ^1}} \right]^2}\right\}.
\label{e4.10}
\end{eqnarray}
Therefore, we can get the following canonical Hamilton's equations,
\begin{eqnarray}
\dot \phi  &=&\frac{{\sqrt { - g} }}{{e\left( {a - {r^2}} \right)}}\left[ {ea{\pi _\phi } + {e^2}a{A_1}
- r\left( {{\partial _1}{\pi ^1}} \right) - er {\phi '} } \right],\nonumber\\
{\dot A_1} &=& \frac{{\sqrt { - g} }}{{{e^2}\left( {{r^2} - a} \right)}}\left[ {{e^2}\left( {{r^2} - a} \right){\pi ^1}
+ {\partial _1}{\partial _1}{\pi ^1} - er{\partial _1}{\pi _\phi } - {e^2}r{\partial _1}{A_1} + e\phi ''} \right], \nonumber \\
{\dot \pi ^1} &=& \frac{{\sqrt { - g} e}}{{a - {r^2}}}\left[ { - a{\pi _\phi } + r\left( {1 + a - {r^2}} \right)  \phi '  } \right]
- \frac{{\sqrt { - g} r}}{{{r^2} - a}}\left( {{\partial _1}{\pi ^1}} \right) \nonumber\\
& &+\frac{{\sqrt { - g} a{e^2}}}{{{r^2} - a}}\left( {1 + a - {r^2}} \right){A_1},\nonumber\\
{{\dot \pi }_\phi }  &=& \sqrt { - g} \left\{ {\left( {\frac{1}{{{r^2} - a}} - 1} \right)\left( er{\partial _1}{A_1} - {\phi ''} \right)
+ \frac{1}{{e\left( {{r^2} - a} \right)}}\left( {er{\partial _1}{\pi _\phi } - {\partial _1}{\partial _1}{\pi ^1}} \right)} \right\}.\label{e4.11}
\end{eqnarray}
For the other two phase space variables, we can easily obtain their equations of motion from the constraints (Eqs.~\eqref{e4.6} and \eqref{e4.7})
with the replacement of the Dirac weak equality by the strong one.

\subsection{Solution}
By eliminating the momenta from the above Hamilton's equations, we have the Euler-Lagrange equations for $\phi$, $A_0$ and $A_1$, respectively,
\begin{eqnarray}
{\partial _0}\left[ {\frac{1}{{\sqrt { - g} }}{\partial _0}\phi  - e\left( {r{A_0} + {A_1}} \right)} \right]
+ \sqrt { - g} {\partial _1}\left[ {e\left( {{A_0} + r{A_1}} \right) - {\partial _1}\phi } \right] &=& 0, \nonumber\\
{\partial _1}\left( {\sqrt { - g} {\partial _1}{A_0} - {\partial _0}{A_1}} \right) + er{\partial _0}\phi  - \sqrt { - g} e{\partial _1}\phi
- \sqrt { - g} {e^2}a{A_0} &=& 0,\nonumber \\
{\partial _0}\left( {\frac{1}{{\sqrt { - g} }}{\partial _0}{A_1} - {\partial _1}{A_0}} \right) + {\partial _0}\phi  - \sqrt { - g} er{\partial _1}\phi
+ \sqrt { - g} {e^2}a{A_1} &=& 0.\label{e4.12}
\end{eqnarray}
After analyzing the three equations of motion with a technique of calculation, we deduce the solutions as follows:
\begin{eqnarray}
\phi  &=& \sigma  - h, \nonumber\\
{A_0} &=& \frac{1}{{ea}}\left[ { \frac{r}{{\sqrt { - g} }}{\partial _0}\sigma  + \left( {a - {r^2}} \right){\partial _1}\sigma
+\frac{{1 - r}}{{\sqrt { - g} }}{\partial _0}h} \right], \nonumber\\
{A_1} &=& \frac{1}{{ea}}\left[ {r{\partial _1}\sigma  + \frac{{a - {r^2}}}{{\sqrt { - g} }}{\partial _0}\sigma
+ \left( {1 - r} \right){\partial _1}h} \right],\label{e4.13}
\end{eqnarray}
where $h$ and $\sigma$ are new variables and satisfy the following equations of motion,
\begin{eqnarray}
\left( {  \frac{{1}}{{\sqrt { - g} }}{\partial _0} - {\partial _1}} \right)h &=& 0,\label{e4.14}\\
\frac{1}{{\sqrt { - g} }}{\partial _0}\left( {\frac{1}{{\sqrt { - g} }}{\partial _0}\sigma } \right) + {\partial _1}{\partial ^1}\sigma
+ {e^2}a\left( {1 - \frac{1}{{{r^2} - a}}} \right)\sigma  &=& 0.\label{e4.15}
\end{eqnarray}
Eq.~\eqref{e4.14} is, as expected, the noncommutative left-moving chiral boson,\citeu{s8}
and Eq.~\eqref{e4.15} is the noncommutative Klein-Gordon equation
describing a free
and massive scalar boson on the extended Minkowski spacetime. The mass is different from that of the model described in section $3$. In addition,
by using Eqs. \eqref{e4.9}, \eqref{e4.11} and \eqref{e4.13}, we calculate the equal-time Dirac brackets for $h$ and $\sigma$,
\begin{eqnarray}
{\left\{ {\sigma \left( x \right),\sigma \left( y \right)} \right\}_{\rm DB}} &=&  0, \nonumber\\
{\left\{ {h\left( x \right),h\left( y \right)} \right\}_{\rm DB}} &=& 0,\nonumber\\
{\left\{ {\sigma \left( x \right),\dot \sigma \left( y \right)} \right\}_{\rm DB}} &=& \frac{{a\sqrt { - g} }}{{\left( {1 + a - {r^2}} \right)
\left( {a - {r^2}} \right)}}\delta \left( {x - y} \right), \nonumber\\
{\left\{ {h\left( x \right),\dot h\left( y \right)} \right\}_{\rm DB}} &=& \frac{{a\sqrt { - g} }}{{1 + a - {r^2}}}\delta \left( {x - y} \right).
\label{e4.16}
\end{eqnarray}

Consequently, we solve completely the noncommutative generalization of the  bosonized GCSM which is depicted by the Lagrangian
Eq.~\eqref{e4.3} and find that the spectrum of the model includes a chiral boson $h(x)$ with the left-chirality and a massive scalar field $\sigma(x)$
with the mass $m^2={e^2}a\left( {{r^2} - a - 1} \right)/\left( {{r^2} - a} \right)$ in the framework of the extended Minkowski spacetime.
In addition, we note that the equations of motion (Eqs.~\eqref{e4.14} and \eqref{e4.15}) and the Dirac brackets (Eq.~\eqref{e4.16})
take their usual forms on the extended Minkowski spacetime
with the $(\tau, x)$ coordinates. This shows the consistency of our generalization and also provides a simple way to solve Eq.~\eqref{e4.15}
as mentioned in the above section.

\subsection{Duality symmetry}
According to the approach utilized in section $3$, we introduce two vector fields $F_\mu$ and $G^\mu$, and write the following parent action
corresponding to Eq.~\eqref{e4.2},
\begin{eqnarray}
S_2^{\rm p}&=& \int {\mathrm{d}t\mathrm{d}x} \left[\frac{1}{{2\sqrt { - g} }}{{\left( {{F_0}} \right)}^2}
- \frac{{\sqrt { - g} }}{2}{{\left( {{F_1}} \right)}^2} + \frac{1}{{2\sqrt { - g} }}{{\left( {{\partial _0}{A_1}} \right)}^2}\right. \nonumber\\
& &+ \frac{{\sqrt { - g} }}{2}{{\left( {{\partial _1}{A_0}}\right)}^2} - {\left( {{\partial _0}{A_1}} \right)}\left({{\partial _1}{A_0}}\right)
- e\left( {r{A_0} + {A_1}} \right){F_0} \nonumber\\
& &\left.+ \sqrt { - g} e\left( {{A_0} + r{A_1}} \right){F_1}+ \frac{{\sqrt { - g} }}{2}{e^2}a\left[ {{{\left( {{A_0}} \right)}^2}
- {{\left( {{A_1}} \right)}^2}} \right]+ {G^\mu }\left( {{F_\mu } - {\partial _\mu }\phi } \right)\right].\label{e4.17}
\end{eqnarray}
The variation of Eq.~\eqref{e4.17} with respect to $G^\mu$ gives $F_\mu = {\partial _\mu }\phi$, which simply shows the equivalence between
Eq.~\eqref{e4.2} and Eq.~\eqref{e4.17}. However, the variation of Eq.~\eqref{e4.17} with respect to $F^\mu$ gives the nontrivial formulas we need,
\begin{eqnarray}
{F_0} &=& -\sqrt { - g} \left[ {  {G_0} - e\left( {r{A_0} + {A_1}} \right)} \right], \nonumber\\
{F_1} &=& e\left( {{A_0} + r{A_1}} \right) - \frac{{{G_1}}}{{\sqrt { - g} }}.\label{e4.18}
\end{eqnarray}
Substituting Eq.~\eqref{e4.18} into the action Eq.~\eqref{e4.17}, we obtain a dual action of the GCSM,
\begin{eqnarray}
{\tilde S}_2^{\rm dual}&=& \int {\mathrm{d}t\mathrm{d}x} \bigg\{  - \frac{{\sqrt { - g} }}{2}{\left( {{G_0}} \right)^2}
+ \frac{1}{{2\sqrt { - g} }}{\left( {{G_1}} \right)^2} + \frac{1}{{2\sqrt { - g} }}{\left( {{\partial _0}{A_1}} \right)^2}
+ \frac{{\sqrt { - g} }}{2}{\left( {{\partial _1}{A_0}} \right)^2} \bigg. \nonumber\\
& & - \left( {{\partial _0}{A_1}} \right)\left( {{\partial _1}{A_0}} \right) + e\sqrt { - g} \left( {r{A_0} + {A_1}} \right){G_0}
+ \frac{{\sqrt { - g} }}{2}{e^2}\left( {a + 1 - {r^2}} \right)\left[ {{{\left( {{A_0}} \right)}^2} - {{\left( {{A_1}} \right)}^2}} \right] \nonumber\\
& &\bigg.- e\left( {{A_0} + r{A_1}} \right){G_1}+ \phi {\partial _\mu }{G^\mu }\bigg\}.\label{e4.19}
\end{eqnarray}
Making the variation of Eq.~\eqref{e4.19} with respect to $\phi$ we have ${\partial _\mu }{G^\mu } = 0$, whose solution is
\begin{eqnarray}
{G^\mu }\left( \varphi  \right) = {\epsilon ^{\mu \nu }}{\partial _\nu }\varphi\equiv{\epsilon ^{\mu \nu }}{F_\nu }(\varphi),\label{e4.20}
\end{eqnarray}
where $\varphi(x)$ is an arbitrary scalar field. After substituting Eq.~\eqref{e4.20} into Eq.~\eqref{e4.19}
we thus get the dual action in terms of $\varphi$,
\begin{eqnarray}
{S}_2^{\rm dual} &=&\int {\mathrm{d}t\mathrm{d}x} \left\{ \frac{1}{{2\sqrt { - g} }}{\left( {{\partial _0}\varphi } \right)^2}
- \frac{{\sqrt { - g} }}{2}{\left( {{\partial _1}\varphi } \right)^2} + \frac{1}{{2\sqrt { - g} }}{\left( {{{\partial _0}{A_1}}} \right)^2}
+ \frac{{\sqrt { - g} }}{2}{\left( {{{\partial _1}{A_0}}} \right)^2} \right. \nonumber\\
& &- \left( {{\partial _0}{A_1}} \right)\left( {{\partial _1}{A_0}} \right)+ e\sqrt { - g} \left( {r{A_0}
+ {A_1}} \right)\left( {{\partial _1}\varphi } \right)- e\left( {{A_0} + r{A_1}} \right)\left( {{\partial _0}\varphi } \right) \nonumber\\
& &\left.+ \frac{{\sqrt { - g} }}{2}{e^2}\left( {a + 1 - {r^2}} \right)\left[ {{{\left( {{A_0}} \right)}^2}
- {{\left( {{A_1}} \right)}^2}} \right]\right\}.\label{e4.21}
\end{eqnarray}

In order to make a comparison between Eq.~\eqref{e4.17} and its dual partner Eq.~\eqref{e4.21}, we introduce three new parameters
\begin{eqnarray}
r' = \frac{1}{r}, \qquad e' = er, \qquad a' = \frac{{a + 1 - {r^2}}}{{{r^2}}},\label{e4.22}
\end{eqnarray}
where $r\neq0$ in general, and rewrite Eq.~\eqref{e4.21} as follows:
\begin{eqnarray}
{S}_2^{\rm dual}& =  &\int {\mathrm{d}t\mathrm{d}x}\left\{\frac{1}{{2\sqrt { - g} }}{\left( {\dot \varphi } \right)^2}
- \frac{{\sqrt { - g} }}{2}{\left( {\varphi' } \right)^2} + \frac{1}{{2\sqrt { - g} }}{\left(  {{\partial _0}{A_1}} \right)^2}
+ \frac{{\sqrt { - g} }}{2}{\left(  {{\partial _1}{A_0}}  \right)^2}
 \right. \nonumber\\
& & - \left( {{\partial _0}{A_1}} \right)\left( {{\partial _1}{A_0}} \right)
- e'{\dot \varphi}  \left( {r'{A_0} + {A_1}} \right)+ \sqrt { - g}e' {\varphi' }\left( {{A_0} + r'{A_1}} \right) \nonumber\\
& &\left.+ \frac{{\sqrt { - g} }}{2}{{e'}^2}a'\left[ {{{\left( {{A_0}} \right)}^2} - {{\left( {{A_1}} \right)}^2}} \right]\right\},\label{e4.23}
\end{eqnarray}
which has the same form as Eq.~\eqref{e4.2} with the replacements of
$\phi$, $r$, $e$ and $a$ by $\varphi$, $r'$, $e'$ and $a'$, respectively. Consequently, we show
that the nocommutative generalization of the GCSM is self-dual with respect to the
dualization of ${G^\mu }\left( \varphi  \right)$ and ${F_\mu }\left( \varphi  \right)$ (Eq.~\eqref{e4.20}) together with the redefinition of
the parameters (Eq.~\eqref{e4.22}).

\section{Gauge invariant GCSM}
After adding the Wess-Zumino term\citeu{s11} to the bosonic action of the GCSM,\citeu{s10}
we write the complete action of the gauge invariant GCSM in the covariant formulation
on the extended Minkowski spacetime with $(\tau, x)$-coordinates,
\begin{eqnarray}
S_3 &=&\int{\mathrm{d}t \mathrm{d}x}\left\{\frac{1}{2}\left( {\partial {}_\mu \phi } \right)\left( {{\partial ^\mu }\phi } \right)
+ e{A^\mu }\left( {{\epsilon _{\mu \nu }} - r{\eta_{\mu \nu }}} \right){\partial ^\nu }\phi +\frac{1}{2}{e^2}a{A_\mu }{A^\mu }
- \frac{1}{4}{F_{\mu \nu }}{F^{\mu \nu }} \right.\nonumber \\
& &\left.+ \frac{1}{2}\left( {a - {r^2}} \right)\left( {\partial {}_\mu \theta }\right)\left( {{\partial ^\mu }\theta } \right)
+e{A^\mu }\left[ {r{\epsilon _{\mu \nu }} + \left( {a - {r^2}} \right){\eta_{\mu \nu }}} \right]{\partial ^\nu }\theta \right\},
\end{eqnarray}
where $\theta(x)$ is the Wess-Zumino field. We rewrite this action in terms of $(t, x)$-coordinates,
\begin{eqnarray}
S_3 &=& \int {\mathrm{d}t\mathrm{d}x} \sqrt { - g} \left\{\frac{1}{2}{{\left( {\frac{{\partial \phi }}{{\partial t}}\frac{1}{{\dot \tau }}} \right)}^2}
- \frac{1}{2}{{\left( {\frac{{\partial \phi }}{{\partial x}}} \right)}^2}
+ \frac{1}{2}{{\left( {\frac{{\partial {A_1}}}{{\partial t}}\frac{1}{{\dot \tau }}} \right)}^2}
+ \frac{1}{2}{{\left( {\frac{{\partial {A_0}}}{{\partial x}}} \right)}^2}
- \frac{{\partial {A_1}}}{{\partial t}}\frac{1}{{\dot \tau }}\frac{{\partial {A_0}}}{{\partial x}}\right.\nonumber \\
& & - e\left( {r{A_0} + {A_1}} \right)\frac{{\partial \phi }}{{\partial t}}\frac{1}{{\dot \tau }}+ e\left( {{A_0} + r{A_1}} \right)
\frac{{\partial \phi }}{{\partial x}} + \frac{1}{2}{e^2}a\left[ {{{\left( {{A_0}} \right)}^2} - {{\left( {{A_1}} \right)}^2}} \right]\nonumber  \\
& &+ \frac{{a - {r^2}}}{2}\left[ {{{\left( {\frac{{\partial \theta }}{{\partial t}}\frac{1}{{\dot \tau }}} \right)}^2}
- {{\left( {\frac{{\partial \theta }}{{\partial x}}} \right)}^2}} \right] + e\left[ {{A_0}\left( {a - {r^2}} \right)- r{A_1}} \right]
\left( {\frac{{\partial \theta }}{{\partial t}}\frac{1}{{\dot \tau }}} \right)\nonumber\\
& &\left.+e\left[ {r{A_0} - {A_1}\left( {a - {r^2}} \right)} \right]\left( {\frac{{\partial \theta }}{{\partial x}}} \right) \right\}, \label{e5.2}
\end{eqnarray}
where the noncommutativity has been encoded into the action through the transformation of coordinates.
Thus we give the Lagrangian,
\begin{eqnarray}
\mathcal{L}_3 & =& \frac{1}{{2\sqrt { - g} }}{(\dot \phi )^2} - \frac{{\sqrt { - g} }}{2}{{\left(  \phi '\right)}^2}
+ \frac{1}{{2\sqrt { - g} }}{( {\dot {A_1}} )^2} + \frac{{\sqrt { - g} }}{2}{{\left(  A_0 ' \right)}^2} - {A_0'} {\dot {A_1}}  \nonumber\\
& &- e{\dot \phi}\left( {r{A_0} + {A_1}} \right)+ \sqrt { - g} e {\phi '}\left( {{A_0} + r{A_1}} \right)
+ \frac{{\sqrt { - g} }}{2}{e^2}a \left[ {{{\left( {{A_0}} \right)}^2} - {{\left( {{A_1}} \right)}^2}} \right] \nonumber\\
& &+ \frac{{a - {r^2}}}{2}\left[ \frac{1}{{\sqrt { - g} }}{{( {\dot \theta })}^2} - \sqrt { - g} {{\left(\theta ' \right)^2}} \right]
+ e{\dot \theta }\left[ {{A_0}\left( {a - {r^2}} \right) - r{A_1}} \right] \nonumber\\
& & + e\sqrt { - g}\theta' \left[ {r{A_0} - {A_1}\left( {a - {r^2}} \right)} \right],  \label{e5.3}
\end{eqnarray}
which is the noncommutative generalization of the gauge invariant GCSM.

\subsection{Equation of motion}
As dealt with in the above two sections, we define the momenta conjugate to $\phi$, $A_0$, $A_1$ and $\theta$, respectively,
\begin{equation*}
\pi _\phi   \equiv \frac{{\partial \mathcal{L}_3}}{{\partial {\dot \phi}}} = \frac{1}{{\sqrt { - g} }}\dot \phi - e\left( {r{A_0} + {A_1}} \right),
\qquad
\pi ^0  \equiv \frac{{\partial \mathcal{L}_3}}{{\partial {\dot A_0 } }} \approx 0,
\end{equation*}
\begin{equation}
\pi ^1  \equiv \frac{{\partial \mathcal{L}_3}}{\partial \dot A_1}= \frac{1}{{\sqrt { - g} }}{\dot A_1 } - {A_0}',  \qquad
{\pi _\theta } \equiv \frac{\mathcal{\partial L}_3}{{\partial  {\dot \theta }}}=
\frac{{a - {r^2}}}{{\sqrt { - g} }}\dot\theta  + e\left[ {\left( {a - {r^2}} \right){A_0} - r{A_1}} \right], \label{e5.4}
\end{equation}
and then give the Hamiltonian through the Legendre transformation,
\begin{eqnarray}
\mathcal{H}_3 &\equiv &\pi ^\mu  \dot A_\mu   + \pi _\phi  \dot \phi  +{\pi _\theta }\dot \theta - \mathcal{L}_3  \nonumber\\
  &=& \sqrt {-g}\left\{\frac{ {e^2}a\left( {1 + a - {r^2}} \right)}{{2\left( {a - {r^2}} \right)}}{\left( {{A_1}} \right)^2}
  + \frac{1}{2}{\left( {{\pi ^1}} \right)^2} -  \left( {{\partial _1}{\pi ^1}} \right){A_0}
  -  e{\phi'}\left( {{A_0} + r{A_1}} \right)\right. \nonumber\\
 & &+ \frac{1}{2}{\left( {{\pi _\phi }} \right)^2} + \frac{1}{2}{\left( {\phi' } \right)^2}
 +  e{\pi _\phi }\left( {r{A_0} + {A_1}} \right) + \frac{1}{{2\left( {a - {r^2}} \right)}}{\left( {{\pi _\theta }} \right)^2}
 + \frac{{a - {r^2}}}{2} {\left( \theta ' \right)^2} \nonumber \\
 & &\left.- \frac{{e{\pi _\theta }}}{{a - {r^2}}}\left[ {\left( {a - {r^2}} \right){A_0} - r{A_1}} \right]
  -  e{\theta '}\left[ {r{A_0} - \left( {a - {r^2}} \right){A_1}} \right]\right\}.
\end{eqnarray}

The definition of momenta ~(Eq.\eqref{e5.4}) provides one primary constraint,
\begin{eqnarray}
\Omega _1  \equiv \pi ^0  \approx 0,
\end{eqnarray}
and its consistency under the time evolution, $\dot{\Omega }_1\approx 0$, gives one secondary constraint,
\begin{equation}
\Omega _2  \equiv  {\partial _1}{\pi ^1} + e{\phi '} - er{\pi _\phi } + e{\pi _\theta } + er{\theta '} \approx 0,
\end{equation}
but no further constraints can be deduced from $\Omega _2$.
As the both constraints are first class, we have to impose two gauge conditions which can be chosen to be ${\partial _\mu }\theta  \approx 0$,
{\em i.e.},
\begin{eqnarray}
\Omega _3  &\equiv& \theta ' \approx 0,\\
\Omega _4  &\equiv& {\pi _\theta } - e\left[ {\left( {a - {r^2}} \right){A_0} - r{A_1}} \right] \approx 0.
\end{eqnarray}
The four constraints therefore constitute a complete set with the non-vanishing equal-time Poisson brackets,
\begin{eqnarray}
\left\{{\Omega_1 \left( x \right),\Omega _4 \left( y \right)} \right\}_{\rm PB} &=& e\left( {a - {r^2}} \right)\delta \left( {x - y} \right),\nonumber\\
\left\{{\Omega_2 \left( x \right),\Omega _3 \left( y \right)} \right\}_{\rm PB} &=& e{\partial _x}\delta \left( {x - y} \right),\nonumber\\
\left\{{\Omega_3 \left( x \right),\Omega _4 \left( y \right)} \right\}_{\rm PB} &=& {\partial _x}\delta \left( {x - y} \right).
\end{eqnarray}
Next, we calculate the inverse elements of the above Poisson brackets and derive the non-vanishing equal-time Dirac brackets,
\begin{eqnarray}
\left\{ {\phi \left( x \right),\pi _\phi  \left( y \right)} \right\}_{\rm DB}  &=& \delta \left( {x - y} \right), \nonumber\\
\left\{ {A_1 \left( x \right),\pi ^1 \left( y \right)} \right\}_{\rm DB}  &=& \delta \left( {x - y} \right).  \label{e5.11}
\end{eqnarray}

Regarding the Dirac weak constraints as strong conditions, we can write the reduced Hamiltonian in terms of the independent variables,
i.e. $\phi$, $\pi_\phi$, $A_1$ and $\pi^1$,
\begin{eqnarray}
{\cal H}_3^r{\rm{ }} &=& {\sqrt { - g} }\left\{\frac{1}{2}\left[ {{{\left( {{\pi ^1}} \right)}^2} + {{\left( {{\pi _\phi }} \right)}^2}
+ {{\left( {\phi '} \right)}^2}} \right] + \frac{1}{{2{e^2}\left( {a - {r^2}} \right)}}{\left( {er{\pi _\phi }
- {\partial _1}{\pi ^1} - e\phi '} \right)^2}\right. \nonumber\\
& &\left.+ \frac{{{e^2}a\left( {1 + a - {r^2}} \right)}}{{2\left( {a - {r^2}} \right)}}{\left( {{A_1}} \right)^2}
+ \frac{{{A_1}}}{{a - {r^2}}}\left[ {ea{\pi _\phi } - r{\partial _1}{\pi ^1} - er\left( {1 + a - {r^2}} \right)\phi '} \right]\right\}.
\end{eqnarray}
Thus we deduce the Hamilton's equations as follows:
\begin{eqnarray}
\dot \phi  &=& \frac{{\sqrt { - g} }}{{e\left( {a - {r^2}} \right)}}\left( {ea{\pi _\phi } - r{\partial _1}{\pi ^1} - er\phi '
+ {e^2}a{A_1}} \right),\nonumber\\
{{\dot \pi }_\phi} &=&\frac{{\sqrt { - g} }}{{e\left( {a - {r^2}} \right)}}\left[ {e\left( {1 + a - {r^2}} \right)\phi ''
- er{\partial _1}{\pi _\phi } + {\partial _1}{\partial _1}{\pi ^1}} \right]
- \frac{{\sqrt { - g} er}}{{a - {r^2}}}\left( {1 + a - {r^2}} \right){\partial _1}{A_1},\nonumber \\
{{\dot A}_1}& =& \sqrt { - g} {\pi ^1} + \frac{{\sqrt { - g} r}}{{a - {r^2}}}{\partial _1}{A_1}
+ \frac{{\sqrt { - g} }}{{{e^2}\left( {a - {r^2}} \right)}}\left( {er{\partial _1}{\pi _\phi } - {\partial _1}{\partial _1}{\pi ^1}
- e\phi ''} \right),\nonumber\\
{{\dot \pi }^1}& =&  \frac{{\sqrt { - g} }}{{a - {r^2}}}\left[ {er\left( {1 + a - {r^2}} \right)\phi '
+ r{\partial _1}{\pi ^1} - ea{\pi _\phi } - {e^2}a\left( {1 + a - {r^2}} \right){A_1}} \right].  \label{e5.13}
\end{eqnarray}
Here we omit the Hamilton's equations for the other four variables of phase space because we can have them easily from the constraints.

\subsection{Solution}
By eliminating the momenta from the above Hamilton's equations, we obtain the corresponding
Euler-Lagrange equations for $\phi$, $\theta$, $A_0$ and $A_1$, respectively,
\begin{eqnarray}
\partial_0\left[ {\frac{1}{{\sqrt { - g} }}{\partial _0}\phi  - e\left( {r{A_0} + {A_1}} \right)} \right]
+ \sqrt { - g} {\partial _1}\bigg[ {e\left( {{A_0} + r{A_1}} \right) - {\partial _1}\phi } \bigg] &=& 0,\nonumber \\
{\partial _0}\left\{ {\frac{{a - {r^2}}}{{\sqrt { - g} }}\left( {{\partial _0}\theta } \right) + e\left[ {\left( {a - {r^2}} \right){A_0}
- r{A_1}} \right]} \right\}+{\partial _1}\bigg\{- \sqrt { - g}  \left( {a - {r^2}} \right)\left( {{\partial _1}\theta } \right)\bigg.  & &\nonumber\\
\bigg.+ e\sqrt { - g} \left[ {r{A_0} - \left( {a - {r^2}} \right){A_1}} \right] \bigg\} &=& 0, \nonumber\\
{\partial _1}\left( {\sqrt { - g} {\partial _1}{A_0} - {\partial _0}{A_1}} \right) +  er{\partial _0}\phi  - \sqrt { - g} e{\partial _1}\phi
- \sqrt { - g} {e^2}a{A_0} & & \nonumber \\
- e\left( {a - {r^2}} \right)\left( {{\partial _0}\theta } \right) - \sqrt { - g} er\left( {{\partial _1}\theta } \right) &=& 0,\nonumber \\
{\partial _0}\left( {\frac{1}{{\sqrt { - g} }}{\partial _0}{A_1}  - {\partial _1}{A_0}} \right)  + e{\partial _0}\phi
- \sqrt { - g} er{\partial _1}\phi+ \sqrt { - g} {e^2}a{A_1} & & \nonumber \\
+ er\left( {{\partial _0}\theta } \right) + \sqrt { - g} e\left( {a - {r^2}} \right)\left( {{\partial _1}\theta } \right) &=& 0.
\end{eqnarray}
Carefully analyzing the above equations of motion with a technique of calculation, we introduce two new variables
$h$ and $\sigma$, and thus have the solutions as follows:
\begin{eqnarray}
\phi  &=& \sigma  - h, \nonumber\\
{A_0} &=& \frac{1}{{ea}}\left[ { \frac{r}{{\sqrt { - g} }}{\partial _0}\sigma  + \left( {a - {r^2}} \right){\partial _1}\sigma
+\frac{{1 - r}}{{\sqrt { - g} }}{\partial _0}h} \right], \nonumber\\
{A_1} &=& \frac{1}{{ea}}\left[ {r{\partial _1}\sigma  + \frac{{a - {r^2}}}{{\sqrt { - g} }}{\partial _0}\sigma
+ \left( {1 - r} \right){\partial _1}h} \right],  \label{e5.15}
\end{eqnarray}
where $h$ and $\sigma$ satisfy the following equations of motion,
\begin{eqnarray}
\left( {  \frac{{1}}{{\sqrt { - g} }}{\partial _0} - {\partial _1}} \right)h &=& 0, \\ \label{e5.16}
\frac{1}{{\sqrt { - g} }}{\partial _0}\left( {\frac{1}{{\sqrt { - g} }}{\partial _0}\sigma } \right) + {\partial _1}{\partial ^1}\sigma
+ {e^2}a\left( {1 - \frac{1}{{{r^2} - a}}} \right)\sigma  &=& 0.  \label{e5.17}
\end{eqnarray}
The first equation describes a noncommutative left-moving chiral boson\citeu{s8} and the second the noncommutative Klein-Gordon equation
for a free and massive scalar boson on the extended Minkowski spacetime.
Moreover, using Eqs.~\eqref{e5.11}, \eqref{e5.13} and \eqref{e5.15} we obtain the equal-time Dirac brackets for the new variables,
\begin{eqnarray}
{\left\{ {\sigma \left( x \right),\sigma \left( y \right)} \right\}_{\rm DB}} &=& 0,\nonumber \\
{\left\{ {h\left( x \right),h\left( y \right)} \right\}_{\rm DB}}&=& 0,\nonumber \\
{\left\{ {\sigma \left( x \right),\dot \sigma \left( y \right)} \right\}_{\rm DB}} &=& {\rm{ }}\frac{{a\sqrt { - g} }}{{\left( {1 + a - {r^2}} \right)
\left( {a - {r^2}} \right)}}\delta \left( {x - y} \right), \nonumber\\
{\left\{ {h\left( x \right),\dot h\left( y \right)} \right\}_{\rm DB}} &=& \frac{{a\sqrt { - g} }}{{1 + a - {r^2}}}\delta \left( {x - y} \right).
\end{eqnarray}

Therefore, we find out the solution of the noncommutative generalization of the gauge invariant GCSM depicted by the Lagrangian
Eq.~\eqref{e5.3}. That is, the spectrum of the model includes a chiral boson $h(x)$ with the left-chirality and a massive scalar field $\sigma(x)$
with the mass $m^2={e^2}a\left( {{r^2} - a - 1} \right)/\left( {{r^2} - a} \right)$ in the framework of the extended Minkowski spacetime.
The above results show that the gauge invariant GCSM has the same spectrum as that of the GCSM, which can be understood easily
because the gauge invariant GCSM under the gauge fixing $\partial_\mu\theta \approx 0$ coincides with the  GCSM.

\subsection{Duality symmetry}
In order to investigate the duality with respect to both $\phi$ and $\theta$, we introduce two pairs of auxiliary vector fields ($F_\mu$, $G^\mu$) and
($P_\mu$, $Q^\mu$), and write the following parent action corresponding to Eq.~\eqref{e5.2},
\begin{eqnarray}
S_3^{\rm p} &=& \int {\mathrm{d}t\mathrm{d}x} \bigg\{\frac{1}{{2\sqrt { - g} }}{{\left( {{F_0}} \right)}^2}
- \frac{{\sqrt { - g} }}{2}{{\left( {{F_1}} \right)}^2} + \frac{1}{{2\sqrt { - g} }}{\left( {{\partial _0}{A_1}} \right)^2}
+ \frac{{\sqrt { - g} }}{2}{\left( {{\partial _1}{A_0}} \right)^2} \bigg.\nonumber\\
& &  - \left( {{\partial _0}{A_1}} \right)\left( {{\partial _1}{A_0}} \right)- e\left( {r{A_0} + {A_1}} \right){F_0}
+ \sqrt { - g} e\left( {{A_0} + r{A_1}} \right){F_1}+ {G^\mu }\left( {{F_\mu } - {\partial _\mu }\phi } \right)\nonumber  \\
& &+ \frac{{\sqrt { - g} }}{2}{e^2}a\left[ {{{\left( {{A_0}} \right)}^2} - {{\left( {{A_1}} \right)}^2}} \right]
+\frac{1}{2}\left( {a - {r^2}} \right)\left[ {\frac{1}{{\sqrt { - g} }}{{\left( {{P_0}} \right)}^2}
- \sqrt { - g} {{\left( {{P_1}} \right)}^2}} \right]  \nonumber\\
& &+ e\left[ {{A_0}\left( {a - {r^2}} \right) - r{A_1}} \right]{P_0} + e\sqrt { - g} \left[ {r{A_0}
- {A_1}\left( {a - {r^2}} \right)} \right]{P_1}\nonumber\\
& &\bigg.+ {Q^\mu }\left( {{P_\mu } - {\partial _\mu }\theta } \right) \bigg\}.\label{e5.19}
\end{eqnarray}
The variations of Eq.~\eqref{e5.19} with respect to $G^\mu$ and $Q^\mu$, respectively, give $F_\mu = {\partial _\mu }\phi$ and
$P_\mu = {\partial _\mu }\theta$,
which simply yields the equivalence between Eq.~\eqref{e5.2} and Eq.~\eqref{e5.19}. However, making the variations of Eq.~\eqref{e5.19}
with respect to $F_\mu$ and $P_\mu$, respectively,  we have
\begin{eqnarray}
{F_0} &=& -\sqrt { - g} \left[ {  {G_0} - e\left( {r{A_0} + {A_1}} \right)} \right],\nonumber \\
{F_1} &=& e\left( {{A_0} + r{A_1}} \right) - \frac{{{G_1}}}{{\sqrt { - g} }},\nonumber\\
{P_0} &=& -\frac{{\sqrt { - g} }}{{a - {r^2}}}\left\{ {  {Q_0} + e\left[ {\left( {a - {r^2}} \right){A_0} - r{A_1}} \right]} \right\}, \nonumber\\
{P_1} &=& \frac{1}{{a - {r^2}}}\left\{ {e\left[ {r{A_0} - \left( {a - {r^2}} \right){A_1}} \right] - \frac{{{Q_1}}}{{\sqrt { - g} }}} \right\}.
\label{e5.20}
\end{eqnarray}
Substituting Eq.~\eqref{e5.20} into the action Eq.~\eqref{e5.19}, we derive a kind of dual versions for the action $S_3$ (Eq.~\eqref{e5.2}),
\begin{eqnarray}
{\tilde S}_3^{\rm dual} &=&\int {\mathrm{d}t\mathrm{d}x}\bigg\{  - \frac{{\sqrt { - g} }}{2}{\left( {{G_0}} \right)^2}
+ \frac{1}{{2\sqrt { - g} }}{\left( {{G_1}} \right)^2} + \frac{1}{{2\sqrt { - g} }}{\left( {{{\partial _0}{A_1}}} \right)^2}
+ \frac{{\sqrt { - g} }}{2}{\left( {{{\partial _1}{A_0}}} \right)^2}\bigg. \nonumber\\
& &- \left( {{\partial _0}{A_1}} \right)\left( {{\partial _1}{A_0}} \right)+ e\sqrt { - g} \left( {r{A_0} + {A_1}} \right){G_0}
- e\left( {{A_0} + r{A_1}} \right){G_1} + \phi {\partial _\mu }{G^\mu }\nonumber\\
& &+ \frac{{\sqrt { - g} }}{2}{e^2}\left( {a + 1 - {r^2}} \right)\left[ {{{\left( {{A_0}} \right)}^2} - {{\left( {{A_1}} \right)}^2}} \right] - \frac{{\sqrt { - g} }}{{2\left( {a - {r^2}} \right)}}{\left( {{Q_0}} \right)^2}  + \frac{{{{\left( {{Q_1}} \right)}^2}}}{{2\left( {a - {r^2}} \right)\sqrt { - g} }}\nonumber\\
& & - \frac{{e\sqrt { - g} }}{{a - {r^2}}}{Q_0}\left[ {\left( {a - {r^2}} \right){A_0} - r{A_1}} \right] 
- \frac{e}{{a - {r^2}}}{Q_1}\left[ {r{A_0} - \left( {a - {r^2}} \right){A_1}} \right] \nonumber\\
& & + \frac{{\sqrt { - g} {e^2}}}{{2\left( {a - {r^2}} \right)}}\left[ {{r^2} - {{\left( {a - {r^2}} \right)}^2}} \right]\left[ {{{\left( {{A_0}} \right)}^2} - {{\left( {{A_1}} \right)}^2}} \right]+ \theta {\partial _\mu }{Q^\mu }\bigg\}.
\label{e5.21}
\end{eqnarray}
Furthermore, we make the variations of Eq.~\eqref{e5.21} with respect to both $\phi$ and $\theta$,
and then obtain the equations ${\partial _\mu }{G^\mu } = 0$
and ${\partial _\mu }{Q^\mu } = 0$, whose solutions are
\begin{eqnarray}
{G^\mu }\left( \varphi  \right) = {\epsilon ^{\mu \nu }}{\partial _\nu }\varphi\equiv{\epsilon ^{\mu \nu }}{F_\nu }(\varphi),\qquad
{Q^\mu }\left( \vartheta  \right) = -{\epsilon ^{\mu \nu }}{\partial _\nu }\vartheta\equiv - {\epsilon ^{\mu \nu }}{P_\nu }(\vartheta),\label{e5.22}
\end{eqnarray}
where $\varphi(x)$ and $\vartheta(x)$ are arbitrary scalar fields. When Eq.~\eqref{e5.22} is substituted into Eq.~\eqref{e5.21},
the dual action is expressed in terms of $\varphi$ and $\vartheta$ as follows:
\begin{eqnarray}
{S_3^{\rm dual}}&=& \int {\mathrm{d}t\mathrm{d}x}\left\{\frac{1}{{2\sqrt { - g} }}{\left( {{\partial _0}\varphi } \right)^2}
- \frac{{\sqrt { - g} }}{2}{\left( {{\partial _1}\varphi } \right)^2} + \frac{1}{{2\sqrt { - g} }}{\left( {{\partial _0}{A_1}} \right)^2}
+ \frac{{\sqrt { - g} }}{2}{\left( {{\partial _1}{A_0}} \right)^2}  \right.\nonumber\\
& & - \left( {{\partial _0}{A_1}} \right)\left( {{\partial _1}{A_0}} \right)+ \frac{{\sqrt { - g} }}{2}{e^2}\left( {a + 1 - {r^2}} \right)\left[ {{{\left( {{A_0}} \right)}^2}
- {{\left( {{A_1}} \right)}^2}} \right]\nonumber \\
& &+ e\sqrt { - g} \left( {r{A_0} + {A_1}} \right)\left( {{\partial _1}\varphi } \right) - e\left( {{A_0} + r{A_1}} \right)\left( {{\partial _0}\varphi } \right)
+ \frac{1}{{a - {r^2}}}\left[ \frac{1}{{2\sqrt { - g} }}\left( {{\partial _0}\vartheta } \right)^2
\right.\nonumber\\
& & - \frac{{\sqrt { - g} }}{2}\left( {{\partial _1}\vartheta } \right)^2 - \frac{{\sqrt { - g} }}{2}{e^2}\left[ {{{\left( {a - {r^2}} \right)}^2}
- {r^2}} \right]\left[ {{{\left( {{A_0}} \right)}^2} - {{\left( {{A_1}} \right)}^2}} \right] \nonumber \\
& &\bigg.\bigg.+ e\left[ {r{A_0} - \left( {a - {r^2}} \right){A_1}} \right] \left( {{\partial _0}\vartheta } \right)+ e\sqrt { - g} \left[ {\left( {a - {r^2}} \right){A_0} - r{A_1}} \right]\left( {{\partial _1}\vartheta} \right)\bigg]\Bigg\}.
\label{e5.23}
\end{eqnarray}

Similar to the case in the above section, we introduce three new parameters
\begin{eqnarray}
r' = \frac{1}{r}, \qquad e' =  er ,\qquad a' = \frac{a}{{\left( {a - {r^2}} \right){r^2}}},\label{e5.24}
\end{eqnarray}
and then rewrite Eq.~\eqref{e5.23} as
\begin{eqnarray}
{S_3^{\rm dual}} &=&\int {\mathrm{d}t \mathrm{d}x}\left\{ \frac{1}{{2\sqrt { - g} }}{\left( {\dot \varphi } \right)^2}
- \frac{{\sqrt { - g} }}{2}{\left( {\varphi '} \right)^2} + \frac{1}{{2\sqrt { - g} }}{\left(  {{\partial _0}{A_1}} \right)^2}
 + \frac{{\sqrt { - g} }}{2}{\left( {{\partial _1}{A_0}} \right)^2}\right. \nonumber\\
& &-\left( {{\partial _0}{A_1}} \right)\left( {{\partial _1}{A_0}} \right)
- e'{\dot \varphi}\left( {r'{A_0} + {A_1}} \right) + \frac{{a' - {r'^2}}}{2}\left[ {\frac{1}{{\sqrt { - g} }}{{( \dot \vartheta )}^2}
- \sqrt { - g} {{( {\vartheta '} )}^2}} \right]  \nonumber\\
& &+ e'{\varphi'}\sqrt { - g} \left( {{A_0} + r'{A_1}} \right)+ e'{\dot \vartheta}\left[ {\left( {a' - {{r'}^2}} \right){A_0} - r'{A_1}} \right]
\nonumber\\
& & \bigg. + \frac{{\sqrt { - g} }}{2}{{e'}^2}a'\left[ {{{\left( {{A_0}} \right)}^2} - {{\left( {{A_1}} \right)}^2}} \right]+ e'{\vartheta '}
\sqrt { - g} \left[ {r'{A_0} - \left( {a' - {{r'}^2}} \right){A_1}} \right]\bigg\},\label{e5.25}
\end{eqnarray}
which has the same form as Eq.~\eqref{e5.2} with the replacements of $\phi$, $\theta$, $r$, $e$ and $a$ by $\varphi$, $\vartheta$, $r'$, $e'$ and $a'$,
respectively. As a result, the noncommutative generalization of the gauge invariant GCSM is self-dual with respect to the dualization of
${G^\mu }\left( \varphi  \right)$ and ${F_\mu }\left( \varphi \right)$
and of ${Q^\mu }\left( \vartheta  \right)$ and ${P_\mu }\left( \vartheta  \right)$
(Eq.~\eqref{e5.22}) together with the redefinition of the parameters (Eq.~\eqref{e5.24}).

\section{Conclusion}
In this paper we briefly review the proposal of the
noncommutative extension of the Minkowski spacetime in which a proper
time is defined in order to connect the $\kappa$-Minkowski spacetime and the extended Minkowski spacetime.
The information of noncommutativity can be encoded from the $\kappa$-Minkowski spacetime into  the extended spacetime. Next, we
apply the proposal to the three models: the interacting model of
Floreanini-Jackiw chiral bosons and gauge fields, the generalized
chiral Schwinger model and its gauge invariant formulation. The noncommutative actions of the three models
are acquired and quantized by the use of the Dirac's method, and then the self-dualities of the actions are
investigated. We find that the self-dualities still remain in the three models, which shows that such a symmetry appears
in a wide context of models related to the interacting theories of chiral bosons and gauge fields.

\section*{Acknowledgments}
Y.-G.M is indebted to the Associate Scheme provided by the Abdus Salam International Centre for
Theoretical Physics where part of the work is performed. This work is supported in part by the National Natural
Science Foundation of China under grant No.11175090, by the Fundamental Research Funds for the Central Universities under grant
No.65030021, and by the Project of Knowledge Innovation Program (PKIP) of Chinese Academy of Sciences under grant No. KJCX2.YW.W10.

\end{document}